\documentclass[10pt,twocolumn,aps,pra,showpacs,superscriptaddress,floatfix,nofootinbib,longbibliography]{revtex4-1}
\usepackage{xcolor}
\usepackage{amsthm}
\usepackage{graphicx}
\usepackage{amstext}
\usepackage{amsmath}            
\usepackage{amssymb}            
\usepackage{graphicx}           
\usepackage{latexsym}
\usepackage{bm}
\usepackage{etoolbox}
\usepackage{breqn}
\usepackage{wrapfig} 
\usepackage{ulem}
\usepackage{float}
\makeatletter
\let\cat@comma@active\@empty
\makeatother
\usepackage{url}

\newcommand{\ket}[1]{| #1 \rangle}
\newcommand{\bra}[1]{\langle #1 |}

\newcommand{\beq}{\begin{eqnarray}}
\newcommand{\eeq}{\end{eqnarray}}

\newcommand{\eq}[1]{Eq.~(\ref{#1})}

\newcommand{\red}[1]{{\color{black} #1}}

\def\II{\mathcal{I}}

\def\<{\langle}
\def\>{\rangle}

\def \info#1{}

\begin{document}

\title{Experimental test of non-macrorealistic cat states in the cloud}

\author{Huan-Yu Ku}
\affiliation{Department of Physics and Center for Quantum Frontiers of Research \& Technology (QFort), National Cheng Kung University, Tainan 701, Taiwan}
\affiliation{Theoretical Quantum Physics Laboratory, RIKEN Cluster for Pioneering Research, Wako-shi, Saitama 351-0198, Japan}

\author{Neill Lambert}
\email{nwlambert@gmail.com}
\affiliation{Theoretical Quantum Physics Laboratory, RIKEN Cluster for Pioneering Research, Wako-shi, Saitama 351-0198, Japan}

\author{Feng-Jui Chan}
\affiliation{Department of Physics and Center for Quantum Frontiers of Research \& Technology (QFort), National Cheng Kung University, Tainan 701, Taiwan}

\author{Clive Emary}
\affiliation{Joint Quantum Centre Durham-Newcastle, School of Mathematics, Statistics and Physics, Newcastle University, Newcastle upon Tyne NE1 7RU, UK}

\author{Yueh-Nan Chen}
\email{yuehnan@mail.ncku.edu.tw}
\affiliation{Department of Physics and Center for Quantum Frontiers of Research \& Technology (QFort), National Cheng Kung University, Tainan 701, Taiwan}
\affiliation{Theoretical Quantum Physics Laboratory, RIKEN Cluster for Pioneering Research, Wako-shi, Saitama 351-0198, Japan}

\author{Franco Nori}
\affiliation{Theoretical Quantum Physics Laboratory, RIKEN Cluster for Pioneering Research, Wako-shi, Saitama 351-0198, Japan}
\affiliation{Department of Physics, The University of Michigan, Ann Arbor, 48109-1040 Michigan, USA}

\date{\today}

\begin{abstract}
The Leggett-Garg inequality attempts to classify experimental outcomes as arising from one of two possible classes of physical theories: those described by macrorealism (which obey our intuition about how the macroscopic classical world behaves), and those that are not (e.g., quantum theory). The development of cloud-based quantum computing devices enables us to explore the limits of macrorealism in new regimes. In particular, here we take advantage of the properties of the programmable nature of the IBM quantum experience to observe the violation of the Leggett-Garg inequality (in the form of a ``quantum witness'') as a function of the number of constituent systems (qubits), while simultaneously maximizing the `disconnectivity', a potential measure of macroscopicity, between constituents. Our results show that two-qubit and four-qubit ``cat states'' (which have large disconnectivity) are seen to violate the inequality, and hence can be classified as non-macrorealistic. In contrast, a six-qubit cat state does not violate the ``quantum-witness'' beyond a so-called clumsy invasive-measurement bound, and thus is compatible with ``clumsy macrorealism". As a comparison, we also consider un-entangled product states with $n=2$, $3$, $4$, and $6$ qubits, in which the disconnectivity is low.

\end{abstract}
\maketitle

\section{Introduction}\label{Int}


The availability of public quantum computers, like the ``IBM quantum experience"~\cite{IBMQ}, promises both applications~\cite{Kandala2017,Wecker14,Balu18,Hsieh2017,Morris19,Mitarai18,Devitt16,Steiger2018,Knill2005,Harper2018} and tests of fundamental physics~\cite{Alsina16,Wang2018,Mooney19}. In particular, as the number of available qubits increases, it potentially allows for a rigorous study of the crossover between classical and quantum worlds~\cite{Preskill2018,Schrodinger35} including tests like the Leggett-Garg inequality (LGI)~\cite{Leggett85,Emary14}.  The LGI was derived as a means to classify experimental outcomes as arising from one of two possible classes of physical theories: those described by macrorealism, and those that are not (e.g., quantum theory).  

A macrorealistic theory is one where the system properties are always well-defined (i.e., obey realism), and in which said properties can be observed in a measurement-independent manner (i.e., measurements just reveal pre-existing properties of the system, and do so in a way that does not change those properties).  Quantum theory obeys neither of these stipulations, but our intuition about the classical world does. Thus, ``macrorealists'' propose that macrorealistic theories apply when the dimension, mass, particle number, or some other indicator of the size of a system is increased, such that the behaviour of suitably-macroscopic systems will tend to obey realism and can be observed without disturbance. 

Over the last 10 years a large variety of experimental tests of the LGI, and its generalizations~\cite{Emary12,Lambert10,Lambert102, Uola18,Kofler13,Halliwell16, Budroni14,Lambert16,Hoffmann_2018,Yueh-Nan14,Shin-Liang16,Bartkiewicz16a,Ku16,Ku18,Li15,Uola18A}, have been performed and violations observed. In typical tests, such as with photon~\cite{Goggin2011,Dressel11,Bartkiewicz16b} and nuclei-electron spin pairs~\cite{Knee2012}, the macroscopicity of the system has been small. And while larger for superconducting qubits~\cite{Huffman17,Knee16,PalaciosLaloy2010} and some of the atom-based examples \cite{Robens15,Budroni15}, testing truly macroscopic systems yet remains a distant goal~\cite{Emary142,Bose18}. 
%

In this work, we take advantage of the programmable nature of the IBM quantum experience to enable tests of increasing macroscopicity by directly increasing the number of constituent parts of the system in a non-trivial way. 
To do so we design a circuit that generates $n$-qubit `cat-state' superpositions of fully polarized configurations, i.e., states which have genuine multipartite entanglement~\cite{Ghne2009} and a large ``disconnectivity", an indicator of macroscopicity~\cite{Leggett85,Emary14,Leggett_2002,Leggett2013,White2016,Dressel2014,Nimmrichter13}. This allows us to see how the violation of the LGI (here in the form of a ``quantum witness''~\cite{CheMing12,Kofler13, Marcus19}) changes as we increase the macroscopicity in terms of the number of constituent qubits. 
%

In addition we augment the basic quantum witness test with a measurement invasiveness test~\cite{Knee16,Wilde2011}, which accounts for ``macroscopically-invasive" measurements by modifying the witness bound.
We term systems which cannot violate the new bound ``clumsy-macrorealistic''.
For the experiments we perform on the IBM quantum experience, our tests show that two-qubit and four-qubit `cat states' clearly violate the quantum witness and are thus non-macrorealistic. On the other hand, as we increase the number of qubits involved in the state to six, the witness value is suppressed, suggesting that this case is compatible with ``clumsy-macrorealism''.
 
Finally, instead of preparing entangled states, we also consider product states with zero entanglement, and hence low disconnectivity (\red{compared with our test using entangled states}), which implies these states are less macroscopic. In comparison with the cat states, we observe that the violation of the witness for these states is more robust to decoherence as the number of qubits is increased. We also show that the quantum witness can serve an additional role as a dimensionality (as in the number of states in the Hilbert space discriminated by the intermediate measurement) witness.

\section{Theoretical background}
We begin with a review of the theoretical background, including definitions of macrorealism, the quantum witness tests,  a description of the cat-states we generate in the device, and an explanation of why they maximize disconnectivity.


\subsection{Quantum witness}\label{Sec_QW}
According to Leggett and Garg,
macrorealistic systems obey two assumptions: macrorealism {\it per se} (MRPS), and non-invasive measurability (NIM)~\cite{Emary14,Leggett85}. MRPS assumes that the system always exists in a definite macroscopic state, and NIM assumes that measurements reveal what that state is, but do not change it.

Under the assumptions of MRPS and NIM, the LGI in the form of a ``quantum witness''~\cite{CheMing12,Kofler13, Marcus19} tells us that if we consider measurements on a system at two times, $t_1$ and $t_2$, the probability of observing outcome $j$ at time $t_2$ should be independent of whether the measurement at the earlier time, $t_1$, was performed or not. This probability is then related to the sum of all joint probabilities in the standard way~\cite{CheMing12}:
\begin{equation}
p_{t_2}(j)=p_{t_2}^{M}(j) = \sum_i p_{t_1}(i)p_{t_2,t_1}(j|i).
\end{equation}
Here, $p_{t_2}(j)$ is the probability of observing the outcome $j$ at time $t_2$, and $p_{t_1}(i)p_{t_2,t_1}(j|i)$ is the joint probability for observing the measurement outcome $i$ at time $t_1$ followed by the outcome $j$ at time $t_2$. 
The superscript $M$ denotes that a measurement was performed at the earlier time $t_1$, and conversely the absence of $M$ implicitly denotes the probabilities are collated from experiments where such an earlier measurement was not performed.

Given these definitions, the quantum witness~\cite{CheMing12} can be defined as the breakdown of the equality $W=0$ where the witness is defined as
\beq
W&=&|p_{t_2}(j) -p_{t_2}^{M}(j)|.\nonumber \\
\label{Eq_quantum_witness}
\eeq
If we find $W\neq 0$, the state at time $t_1$ is said to be non-macrorealistic, 
 in the sense that the assumptions \red{ of either MRPS or NIM (or both)} are shown to be invalid for it.

The assumption of NIM is hard to 
justify, even if we assume MRPS 
holds. We can modify Eq.~\eqref{Eq_quantum_witness} to take into account certain types of invasive measurements by allowing the measurement process at time $t_1$ to change the macroscopic state of the system. In this case, the relationship between marginal and joint probabilities can be extended to~\cite{CheMing12,Oreshkov2012}
\beq
p_{t_2}^{M}(j) = \sum_k p_{t_2,t_1}(j|k) \sum_i \epsilon^M(k|i) p_{t_1}(i),
\label{Eq_clumsy_measurement}
\eeq
which incorporates the probability $\epsilon^M(k|i)$ that observing the system in state $i$ at $t_1$ can cause the system to change to state $k$.  We dub this the assumption of  ``macroscopically-invasive measurements''.

As shown in Appendix~\ref{App_measurement_invasiveness_test}, combining this assumption with Eq.~\eqref{Eq_quantum_witness} gives us an inequality for ``clumsy-macrorealism''
\beq
  W \leq \max_{i} [\II(i)]; \qquad
  \II(i)=|1-\epsilon^M(i|i)|.
\label{Eq_disturbance_2}
\eeq
We call $\II(i)$ an invasiveness test, and it can be evaluated in an additional experimental run by preparing the system in state $i$, performing a measurement, and checking whether it is still in that state immediately after said measurement. If we observe that the inequality in Eq.~\eqref{Eq_disturbance_2} is violated we can say that the both macrorealism and non-clumsy-macrorealism do not hold. 

\red{Under the most clumsy of measurements, the clumsy-macrorealism bound can be unity if $\epsilon^M(i|i)=0$; which occurs when the measurement so strongly disturbs the system the given state $i$ is completely changed into some other states $j\neq i$. Therefore, we note that the quantum witness and the measurement invasiveness test should be implemented under the same conditions. }

\subsection{Witness violations with cat states}\label{macro}
One of the goals of the Leggett-Garg inequality is to identify whether the macroscopic nature of a given system influences whether it behaves in a ``quantum  way'' or in a macrorealistic fashion. While definitions of macroscopicity are myriad~\cite{Nimmrichter13}, Leggett suggested that a minimal starting point are the \textit{extensive difference} and the \textit{disconnectivity}~\cite{Leggett_2002,Leggett2013}. The former compares the difference in magnitude of the observable outcomes to some fundamental physical scale. Recent experiments have attempted to maximize the extensive difference with a macroscopically large superconducting flux qubit~\cite{Knee16}.

The disconnectivity, in contrast, arises from considering that a violation of the witness by a quantum system arises because, at time $t_2$, quantum dynamics can generate superpositions of ``macroscopic states''. Simultaneously, these superpositions are collapsed by measurement, and hence LGI and quantum witness tests are violated.  Thus, if an object is composed of many `particles',  we want the macroscopic nature of the system to contribute to the `superposition of macroscopic states' in a non-trivial way, i.e.,  a large number of the particles should have different states in the `branches' of the superposition. For instance, a Bell state $1/\sqrt{2}(\ket{11}+\ket{00})$ satisfies the above statement with both qubits having clearly different states in the two branches of the superposition, while the product state $1/\sqrt{2} (\ket{0} + \ket{1}) \otimes (\ket{0} + \ket{1})$ does not. This recalls the idea that in a Schr\"odinger's cat thought experiment, the whole cat is in superposition, not just one whisker.  

To put such a definition in a quantitative format, Leggett argued~\cite{Leggett_2002,Leggett2013} that the disconnectivity can be defined as the \red{``number"} of correlations (between constituents) one needs to measure to distinguish a linear superposition \red{between two branches} from a mixture (which are indistiguishable with single-particle measurements alone). A potential quantitative measure proposed by Leggett in~\cite{Leggett2013} is as follows: considering $n$ spins, for any integer $n' \leq n$, the reduced von-Neumann entropy \red{(also known as the entanglement entropy, a measure of entanglement for bi-partite pure states)} of the state $\rho_{n'}$ (having traced out the other spins) is $S_{n'} = -\mathrm{Tr} \rho_{n'} \mathrm{ln} \rho_{n'}$. Leggett then defined the disconnectivity $\Gamma$ as the maximum value of $n'$ such that
\begin{equation}
\begin{aligned}
\delta_{n'}&=\frac{S_{n'}}  {\mathrm{min}_m (S_m + S_{n'-m})}< \eta,
\end{aligned}
\label{Eq_disconnectiviy}
\end{equation}
\red{where $\eta$ is a small value that sets the bound between classical mixtures and entangled states (see below).}
Here one assigns $\delta_{n'}=1$ when $S_{n'}=\mathrm{min}_m (S_m + S_{n'- m})=0$ \red{and defines $\delta_1=0$}. With this definition one can see that states which are `globally' pure but locally mixed give large values of disconnectivity, implying the mixed-ness arises from global entanglement. 

\red{Considering an $n$-body pure entangled state like the GHZ state we use in our experiment, we will have a vanishing numerator and non-vanishing denominator,  leading to $\delta_{n}=0$ and thus $\Gamma=n$. On the other hand, for a product state, or a mixture of product states, one finds $\delta_2 = 1$ and $0.5$ respectively, and hence $\Gamma=1$ for these cases. (As an aside, this suggests a possible choice of $\eta =0.5$ as a bound to delineate between mixtures of product states and mixed entangled states in Eq.~\eqref{Eq_disconnectiviy}). } It is clear that the disconnectivity is strongly related to definitions of genuinely multipartite \red{pure-state} entanglement, a connection which is discussed in-depth in~\cite{Leggett_2002,Leggett2013}. 

\red{In the tests of macrorealism performed to date, most are arguably in the regime of $\Gamma=1$; particularly those employing single photons, electron or nuclear spins, and so on \cite{Goggin2011,Dressel11,Bartkiewicz16b, Knee2012}.  On the other hand, the question of the disconnectivity of a single superconducting qubit \cite{Huffman17,Knee16,PalaciosLaloy2010} has been open to debate (see the supplementary information of \cite{Knee16} for an in-depth discussion).  Our approach here, irrespective of the disconnectivity of the constituent qubit, provides a way to increase the overall disconnectivity by constructing large cat-states of many entangled qubits.}

  To translate this onto the IBM quantum experience, we identify the macroscopic states $i$ of Sec.~\ref{Sec_QW} with the $n$-qubit computational basis states of the quantum register, as revealed by standard read-out measurements. We denote, where appropriate,  these macroscopic states with classical bit-strings, such that what would be  $\ket{0}^{\otimes n}$ in the bra-ket notation we write as $\{0\}^n$.

\begin{figure}[tbp]
\includegraphics[width=1\columnwidth]{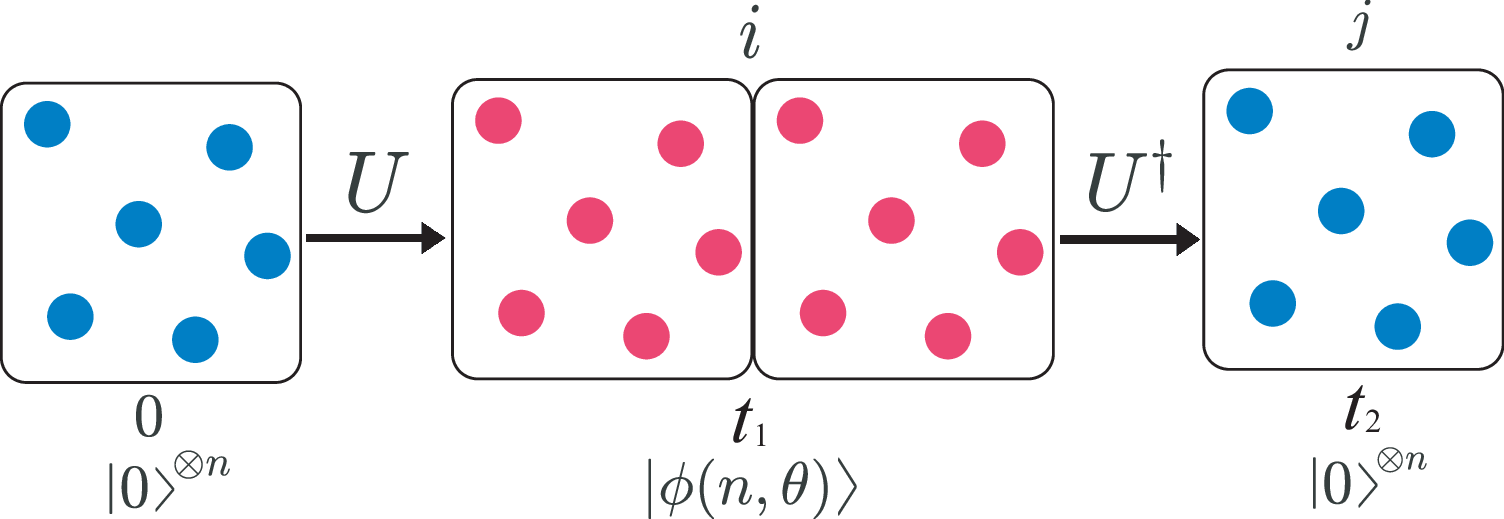}
\caption{We prepare $n$ qubits on the state $\ket{0}^{\otimes n}$ (blue) at time $t_0$. A unitary $U$ transfers the system into the entangled cat state $\ket{\phi(n,\theta)}=\cos{\frac{\theta}{2}}\ket{0}^{\otimes n}+\sin{\frac{\theta}{2}}\ket{1}^{\otimes n}$ (red) at time $t_1$. Then, an inverse unitary $U^{\dagger}$ is performed to the entangled system, such that the system returns back to the state $\ket{0}^{\otimes n}$ at time $t_2$.
The outcomes $i$ and $j$ are obtained at $t_1$ and $t_2$, respectively.
}
\label{system} %
\end{figure}

  To generate superposition states with high disconnectivity, we design circuits in the IBM quantum experience to produce an evolution which starts with all qubits in the product state $\ket{0}^{\otimes n}$ at time $t_0$, and then ideally implements a unitary $U(n,\theta)$ that creates an entangled $n$-qubit `cat state' at `time $t_1$', namely
  \begin{equation}
  \rho_{t_1}=\ket{\phi(n,\theta)}\bra{\phi(n,\theta)},
  \label{Eq_cat_state}
  \end{equation}
  where $\ket{\phi(n,\theta)}=\cos{\frac{\theta}{2}}\ket{0}^{\otimes n}+\sin{\frac{\theta}{2}}\ket{1}^{\otimes n}$, with real coefficient $\theta$ (which for $\theta =\pi/2$ and $n>2$ are GHZ states).  According to the witness prescription, a measurement is then either performed or not performed on this state. We then choose the evolution for $t_1$ to $t_2$ to be given by the inverse unitary transformation $U^{\dagger}$ so that, in the situation where no intermediate measurement has been performed, the entangled state is, ideally,  `evolved back' to the starting state and $\rho_{t_2}=\ket{0}\bra{0}^{\otimes n}$ (see Fig.~\ref{system} for a schematic description).  In the witness itself we choose to only look at the probability of being in that particular macroscopic state, i.e., $j  \equiv \{0\}^{n}$ at $t_2$.
  

In an ideal quantum system, undergoing evolution described by Eq.~\eqref{Eq_cat_state}, one can trivially calculate that $p_{t_2}(\{0\}^n)=1$, $p_{t_1}(\{0\}^n)=\cos^2{\left(\frac{\theta}{2}\right)}$, $p_{t_1}(\{1\}^n)=\sin^2{\left(\frac{\theta}{2}\right)}$, $p_{t_2,t_1}(\{0\}^n|\{0\}^n)=\cos^2{\left(\frac{\theta}{2}\right)}$ and $p_{t_2,t_1}(\{0\}^n|\{1\}^n)=\sin^2{\left(\frac{\theta}{2}\right)}$ (see also Appendix \ref{Appendix quantum circuit direct measure}).
The corresponding quantum witness evaluates as
\begin{equation}
W=1-\cos^4\!{\left(\frac{\theta}{2}\right)}-\sin^4\!{\left(\frac{\theta}{2}\right)}.
\label{Eq_analytic_W}
\end{equation}
Equation~\eqref{Eq_analytic_W} shows that the quantum witness in Eq.~\eqref{Eq_quantum_witness} is violated  for any $\theta \neq k \pi$ with $k=0,1,2,\ldots$.  

\subsection{Witness violations with product states}\label{prod}
It is interesting to compare the situation described in Eq.~\eqref{Eq_cat_state} to an example which still uses many qubits but has low disconnectivity. In this case, the maximally entangled states we used previously are now replaced by a product of single-qubit superposition states at time $t_1$, 
\begin{equation}
\ket{\psi(n)}=\frac{1}{\sqrt{2^n}}\left(\ket{0}+\ket{1}\right)^{\otimes n},
\label{Eq_prod}
\end{equation} 
which has the lowest disconnectivity of $\Gamma=1$~\cite{Leggett_2002,Leggett2013}.
Surprisingly, this product state saturates the maximum quantum bound of the witness, 
which is given by \cite{Greg15}
\begin{equation}
  W_{\text{max}}=1-\frac{1}{D_{\text{Ideal}}},
\label{Eq_W_max}
\end{equation}
where $D_{\text{Ideal}}$ is the number of states spanning the Hilbert space, obtained when the intermediate measurement process discriminates all $D_{\mathrm{Ideal}}$ states. \red{We note that this bound is derived  under the assumptions of quantum mechanics (not clumsy-macrorealism).}
\red{As an aside, we mention that combining Eq.~\eqref{Eq_W_max} with the clumsy-macrorealistic bound in Eq.~\eqref{Eq_disturbance_2}, we can obtain that we need $\epsilon^M(i|i)\geq [D_{\text{Ideal}}]^{-1}$ for the possibility of our system to violate clumsy-macrorealism at all.}

In our experiments, since we individually measure every qubit in the computational basis, $D_{\text{Ideal}}=2^n$ with $n$ number of qubits. Because of this relation between the maximum violation and the dimensionality of the states contributing to the quantum witness, a secondary application as a dimensionality witness arises. In our previous example of cat states, even though we had many qubits, and high disconnectivity, the effective dimension of the states involved in the test of the witness was low, because it was dominated by just two states, $\ket{0}^{\otimes n}$ and $\ket{1}^{\otimes n}$.

\section{Circuit implementation}\label{Sec_Circuit_implementation} 

We use the processor \rm{IBM Q$5$ Tenerife} to experimentally test the $n=2$ cat state  with $\theta\in\{0,\pi/8,2\pi/8,3\pi/8,4\pi/8\}$. With the $14$-qubit processor \rm{IBM Q$14$ Melbourne}, the GHZ states are implemented by considering $\theta=\pi/2$ for $n=4,$ and $6$. The IBM quantum experience only allows for a single measurement to be performed on each qubit. This makes collating the two-time correlation functions required by the quantum witness difficult. To overcome this restriction, we use CNOT gates between `system qubits' and `ancilla qubits', followed by measurements on the ancilla qubits, to perform the intermediate measurement. As a consequence, we are restricted to a maximal qubit number of $6$, with correspondingly $6$-ancilla qubits for measurements at time $t_1$.

From the initial state $\ket{0}^{\otimes n}$, ``cat states'' can be 
obtained by performing the unitary transformation $U$. The unitary $U$ can be decomposed into several parts. It contains the following single-qubit operation applied to the first qubit:
\begin{equation}
U_3(\lambda,\vartheta,\theta)=\left(
\begin{array}{ccc}
\cos{\theta/2} & -e^{i\lambda}\sin{\theta/2} \\
e^{i\vartheta}\sin{\theta/2} & e^{i(\lambda+\vartheta)}\cos{\theta/2}
\end{array}
\right),
\label{Eq_U3}
\end{equation}
with $\lambda=\vartheta=0$, followed subsequently by a series of CNOT gates between the first qubit and each of the others in turn.
The inverse operation $U^{\dagger}$ is given by applying CNOT gates \red{again before} applying the $U^{\dagger}_3=U_3(0,0,-\theta)$ gate on the first qubit. Appendix~\ref{Appendix quantum circuit direct measure} presents a detailed schematic example for a two-qubit system as well as the details of how these operations function.

To generate the product of superposition states described in Eq.~\eqref{Eq_prod}, we perform Hadamard gates on each qubit individually at $t_1$.
At time $t_2$, the Hadamard gates, which are self-inverse, are performed to again obtain a state $\ket{0}^{\otimes n}$. 

\section{Noise simulation}\label{Sec Noise simulation}

Before discussing the experimental results, we introduce a numerical noise model which will assist in understanding two of the main experimental features: suppression of the witness violation due to dephasing, and accidental enhancement of the witness violation due to macroscopically-invasive measurements. 

In the following, to include the influence of decoherence and gate infidelities in a simulation of the quantum circuit, we consider a simple strategy where we assume that each gate is performed perfectly, and instantaneously, after which follows a period of noisy evolution for the prescribed gate time (which can be substantial for two-qubit gates). 
  During such periods,
the dynamics of the system can be described by the following Lindblad master equation~\cite{Johansson2012,Johansson2013}:
\begin{equation}
\begin{aligned}
\dot{\rho_{\text{s}}} &= \sum_i^N \frac{\red{\gamma_{T_1}^i}}{2}\left[ 2\sigma_{-}^{i}\rho_{\text{s}}(t)\sigma_{+}^{i} - \sigma_{+}^{i}\sigma_{-}^{i}\rho_{\text{s}}(t) - \rho_{\text{s}}(t)\sigma_{+}^{i}\sigma_{-}^{i} \right]\\
&+ \sum_i^N \frac{\red{\gamma_{T_2}^i}}{2}\left[ 2\sigma_{z}^{i}\rho_{\text{s}}(t)\sigma_{z}^{i} - {\sigma_{z}^{i}}^2\rho_{\text{s}}(t) - \rho_{\text{s}}(t){\sigma_{z}^{i}}^2 \right].
\end{aligned}
\label{Eq_master_D}
\end{equation}
where $\sigma_{+}^{i}$, $\sigma_{-}^{i}$, $\sigma_{z}^{i}$ represent the raising, lowering and Pauli-$Z$ operators of the $i$th qubit, respectively. \red{Here, we consider the  coefficients to be uniformly $\gamma_{T_1}^i=1/T_1$ and $\gamma_{T_2}^i=[T_2^{-1} - (2T_1)^{-1}]/2~\forall i$.}
The parameters above (like the energy relaxation time $T_1$, dephasing time $T_2$, and gate times) are publicly available and can also be reconstructed by the user using well-defined protocols in the IBM quantum experience.

Equation~\eqref{Eq_master_D} can be derived by assuming that the influence of the environment obeys the standard Born-Markov-Secular approximations. The first and second lines respectively describe energy dissipation and pure dephasing. 
For the comparison to the experimental data we use values for $\gamma_{T_1}$ and $\gamma_{T_2}$ which approximately fit the order of magnitude of the published data ($T_1=46~\mu$s and $T_2=13.5~\mu$s) such that the numerical simulations approximate the observed experimental results well.

Although in general the noise suppresses the violation of the quantum witness, we will show later that the experimental result with $\theta=0$, in which the state has no superposition, is not exactly $0$ as one might expect. From our numerical simulation we find that this non-zero value is due to imperfect gate operations; in particular the ancilla CNOT gates for the intermediate measurement.
For example, during the intermediate measurement step, for $\theta=0$ we ideally expect the system to remain in the $\ket{0}^{\otimes n}$ state for the whole duration of the experiment. However, during the intermediate measurement state the system may be accidentally excited from state $\ket{0}^{\otimes n}$ into other states. This is precisely a ``clumsy'' macroscopically-invasive measurement, {such that $W\neq 0$, even though the state of the system obeys MRPS}. With the effective identity operations $U=U^{\dagger}=\openone$ in $\theta=0$ case, this example exactly corresponds to the measurement invasiveness test we described in the previous section. 

In our minimal simulation, we simply model the effects of such errors by the following extra Lindblad terms for all qubits $i$:\red{
\begin{equation}\label{gamerr}
\frac{\gamma^{i}_{\text{Errors}}}{2}\left[ 2\sigma_{+}^{i}\rho_{\text{s}}(t)\sigma_{-}^{i} - \sigma_{-}^{i}\sigma_{+}^{i}\rho_{\text{s}}(t) - \rho_{\text{s}}(t)\sigma_{-}^{i}\sigma_{+}^{i} \right],
\end{equation}
where $\gamma^{i}_{\text{Errors}}$ is the coefficient to simulate the gate errors for each qubit (which again we take to be uniform $\gamma^{i}_{\text{Errors}}=\gamma_{\text{Errors}}$).} This noise term is phenomenological, and is introduced to capture the noisy imperfect nature of the intermediate measurements which can cause an effective excitation of the qubits, as described above. For our quantum circuits, we determine this value ($\gamma_{\text{Errors}}=8.5\times 10^{-2}~ \mu \text{s}^{-1}$). 

\red{While we primarily use this parameter to fit the witness violation, we point out that when $\theta=0$, the circuit implementation of the quantum witness is identical to the clumsy-measurement test because the intermediate state is simply $\{0\}^2$ in the computational basis. Thus, the noisy simulation of the parameter $\gamma_{\text{Errors}}$ estimates not only the quantum witness but also the measurement invasiveness in that round of the experiment (see Fig.~\ref{Fig_result_for_small_cat}).}

\section{Experimental results}\label{Experimental_results}
We now present the experimental results from the IBM quantum experience, including the measurement invasiveness test and the value of the quantum witness for both the cat states and the product states.
\subsection{Measurement invasiveness of the IBM quantum experience}\label{Sec_disturbance}

\red{To evaluate our proposed modified bound on the witness, we must run additional experiments wherein one prepares and measures all possible quantities $\II(i)$. 
Although we do not explicitly test the quantum witness for a single-qubit, we do test the measurement invasiveness for this case, to check the effect of the clumsy measurement in the IBM quantum experience. The average and maximum values of the invasiveness of single, two, four, and six qubits  are shown in Table~\ref{table_disturbance_two}. While we only present the invasiveness of the state $\left\{0\right\}^n$, as it usually presents the largest disturbance, we do prepare ``all possible states" in the computational basis for single, two, four-qubit systems, in order to test the invasiveness, see Appendix~\ref{Appendix_invasiness_four}.

In general, as we increase the number $n$ of qubits involved in the experiment, testing the invasiveness can be challenging because there are a total of $2^{n}$ circuits to be generated to prepare all possible states $i$. However, if one finds a $\II(i)$ which is already greater than the observed witness violation, it is of course unnecessary to continue. For example, for the six-qubit case, instead of preparing all possible macroscopic states $i$, we only consider the state $\left\{0\right\}^6$ because the experimental value of the quantum witness we observe for the six-qubit case later is already lower than the invasiveness quantity:
\begin{equation}
\begin{aligned}
\II(\left\{0\right\}^6)&=|\;1-\epsilon^M(\left\{0\right\}^6|\left\{0\right\}^6)\;|.\\
\end{aligned}
\label{Eq_invasiveness_bound}
\end{equation}

In all cases ($n=1,2,4,6$), we test the measurement invasiveness in $25$ different experiments across different days, each consisting of $8,192$ runs. This was done because the tests were not performed at the same time as the data collection for testing the actual witness itself. On these long timescales, the IBM quantum experience exhibits fluctuations in parameters, including coherence times, gate fidelities, etc, and thus we introduce a variance in this test that represents these fluctuations. We note that the results of single, and two-qubit systems are obtained from \rm{the IBM Q$5$ Tenerife}, while the results for four, and six-qubit systems are obtained from \rm{the IBM Q$14$ Melbourne}.}


\begin{table}[!htbp]
\centering  
\begin{tabular}{|c|c|c|} \hline
$n$&$\II_{\text{Max}}(\left\{0\right\}^n)$&$\II_{\text{Ave}}(\left\{0\right\}^n)$\\ \hline
$1$&$0.023\pm 0.004$ & $0.016\pm 0.003$ \\ \hline
$2$&$0.077\pm 0.008$ & $0.068\pm 0.006$ \\ \hline
$4$&$0.146\pm 0.006$ & $0.106\pm 0.019$ \\ \hline 
$6$ &$0.686\pm 0.007$ &$0.310\pm 0.225$ \\ \hline
\end{tabular}
\caption{Table of the measurement invasiveness parameters of states $\left\{0\right\}^n$ for single, two, four, and six-qubit systems, respectively. \red{The results of all possible states $i$ in the computational basis for single, two, and four-qubit cases are given in Appendix~\ref{Appendix_invasiness_four}.} Here we perform $25$ experiments, across multiple days, to take into account the variability in the IBM quantum system parameters. Each experiment consists of $8,192$ runs. The maximal and average values,  over experiments,  of the measurement invasiveness are obtained from \rm{the IBM Q$5$ Tenerife} for the single and two-qubit systems, while the results for four, and six-qubit systems are from \rm{the IBM Q$14$ Melbourne}. The simulation results, using the noise model described in the main text, are respectively $\II_{\text{Sim}}(\left\{0\right\}^1) = 0.031$ and $\II_{\text{Sim}}(\left\{0\right\}^2) = 0.061$ for the single and two-qubit cases.}
\label{table_disturbance_two}
\end{table}

\subsection{Maximally entangled state with high disconnectivity}
\begin{figure}[tbp]
\includegraphics[width=1\columnwidth]{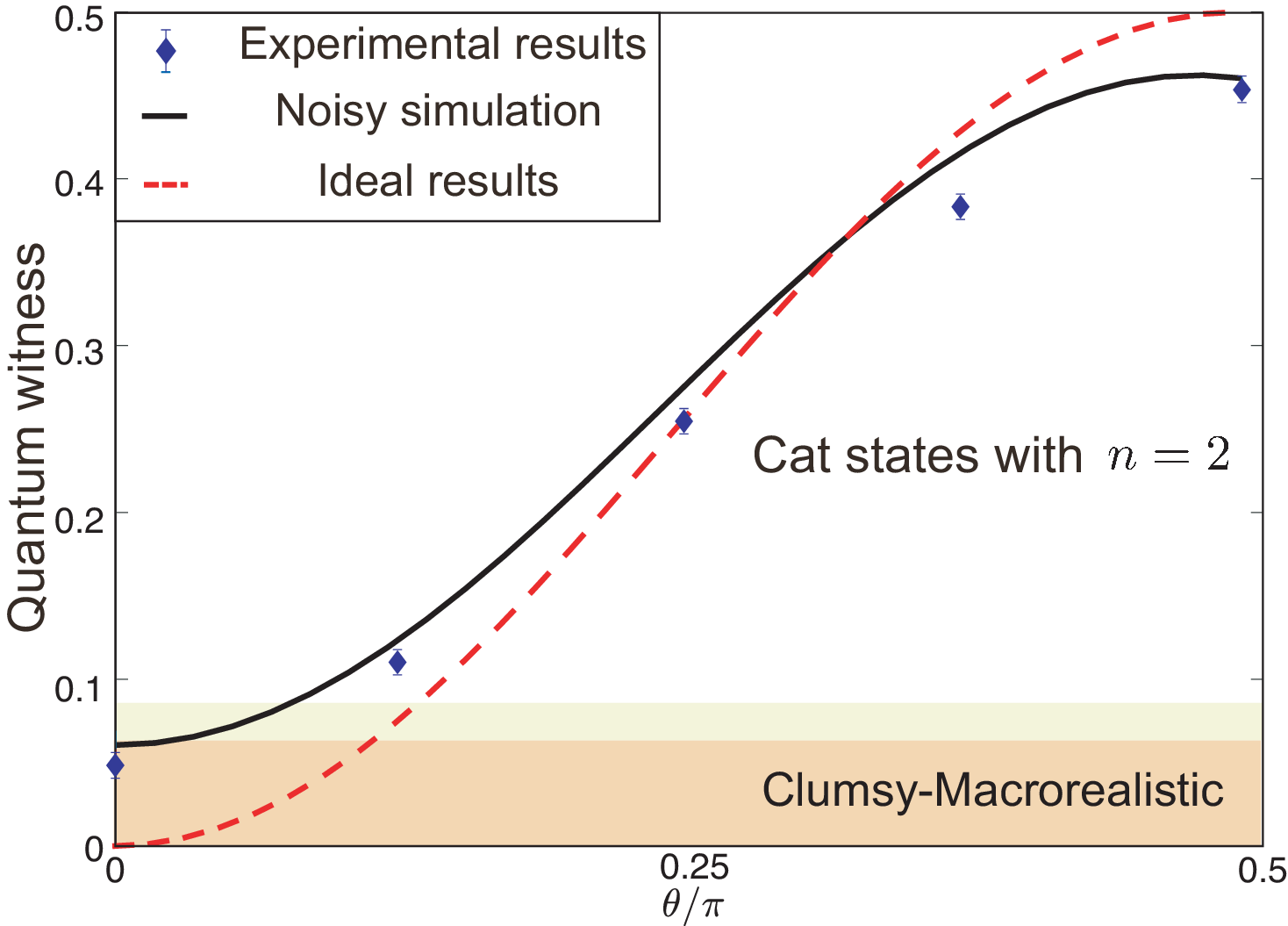}
\caption{The value of the quantum witness of $n=2$ cat states. The circuit is designed to produce the state $\ket{\phi(2,\theta)}=\cos{\frac{\theta}{2}}\ket{0}^{\otimes 2}+\sin{\frac{\theta}{2}}\ket{1}^{\otimes 2}$ at an intermediate time. The experimental results obtained by the IBM quantum experience are shown by blue diamonds. The theoretical results, with and without noise simulation, are shown by red dashed and black solid curves, respectively. Obviously, the quantum witness increases with the parameter $\theta$, but also shows a residual violation due to the macroscopically-invasive measurements backaction and gate error at $\theta =0$.
We simulate the influence of decoherence and gate infidelities by Lindblad-form master equations Eq.~\eqref{Eq_master_D} and Eq.~\eqref{gamerr}. The coefficients of relaxation time $T_1=46~\mu$s, dephasing time $T_2=13.5~\mu$s, and gate-error coefficient $\gamma_{\text{Errors}}=8.5\times 10^{-2}~\mu \text{s}^{-1}$ are determined by approximately fitting the experimental results. The gray and orange shaded areas at the bottom are the clumsy-macrorealistic regimes determined by the maximal and average invasiveness tests in Table~\ref{table_disturbance_two}.
Note that the invasiveness test including the standard deviation does not depend on $\theta$. The experimental uncertainties are derived from the multinomial distribution and error propagation except for the average disturbance case which is the variance across $25$ repeated experiments.}
\label{Fig_result_for_small_cat} %
\end{figure}

Figure~\ref{Fig_result_for_small_cat} shows experimental data for the $n=2$ cat state with $\theta\in\{0,\pi/8,2\pi/8,3\pi/8,4\pi/8\}$. We also show the theoretical predictions both with and without noise simulation, as well as the modified witness bound based on the measurement invasiveness tests of Sec.~\ref{Sec_disturbance}. From Fig.~\ref{Fig_result_for_small_cat} we observe that the maximum value of the quantum witness occurs when the entanglement parameter $\theta=\frac{\pi}{2}$, which is the maximally entangled state. 

At $\theta = 0$, we find that the value of the quantum witness is lower than the average experimental measurement invasiveness tests from Sec.~\ref{Sec_disturbance}, implying it is consistent with macrorealism. Interestingly, there is a residual small violation of the witness even though this is not predicted by the simple `pure states' \red{$\{0\}^2$} expression in Eq.~\eqref{Eq_analytic_W}. This `invasiveness' represents a classically invasive measurement. For example, in our simulation plotted in Fig.~\ref{Fig_result_for_small_cat}, we observe that the $\theta=0$ non-zero witness value arises directly from $\gamma_{\mathrm{Errors}}$ in~\eq{gamerr} (i.e., if we set $\gamma_{\mathrm{Errors}}=0$ the witness value in the simulation falls to zero). \red{Thus, as discussed in the simulation section, the $\gamma_{\mathrm{Errors}}$ is related to the clumsy measurability $\epsilon^M({0}^2|{0}^2)$ since at $\theta=0$ the intermediate state corresponds to $\{0\}^2$.}


When we consider the results of the $n=4$ and $n=6$ cat states in Table.~\ref{table_big_cat}, we see that the magnitude of the quantum witness violation drastically decreases, as compared to the $n=2$ case. The value of the quantum witness for the four-qubit cat state is still larger than the invasiveness test, implying a non-macrorealistic behavior. In contrast, while the invasiveness test for six qubits is only performed with the $\left\{0\right\}^6$ state, we see that the six-qubit cat state does not exceed these tests. Thus, we can conclude the six-qubit system is compatible with a clumsy-macrorealistic description. This result shows that the IBM quantum experience tends to a clumsy-macroscopic realistic behaviour as the number of qubits increases, due to the increased influence of decoherence and dephasing processes as the circuit complexity, or ``depth''~\cite{Preskill2018}, increases.

\subsection{Product states with low disconnectivity}

\begin{table}[!]
\centering  
\begin{tabular}{|c|c|c|} \hline
  & $n=4$ & $n=6$   \\ \hline
$W$ & $0.245\pm 0.008$ & $0.181\pm 0.007$   \\ \hline
$\II_{\text{Max}}(\left\{0\right\}^n)$& $0.146\pm 0.006$ & $0.686\pm 0.007$ \\ \hline
Relative difference& $0.099\pm 0.010$ & $-0.505\pm 0.010$\\ \hline
\end{tabular}
\caption{Table of the quantum witness $W$, maximal measurement invasiveness $\II_{\text{Max}}(\left\{0\right\}^n)$, and their relative difference for the larger cat states $n=4$ and $n=6$.}
\label{table_big_cat}
\end{table}


The values we observe for the quantum witness with the product states defined in \eqref{prod} are shown in Table~\ref{table_superposition states}. We find that, compared to cat states, the product states witness values are more robust as we increase $n$, reflecting the increased sensitivity of cat states used in the previous section to dephasing and decoherence, and the lower circuit depth (and hence less time being spent exposed to noise) of this product state example. 
Due to inevitable noise described above, the observed values of the quantum witness for each $n$ do not reach precisely the corresponding theoretically predicted maximum possible values of $W_{\text{max}} = 1-D^{-1}_{\text{Ideal}}$ with $D_{\text{Ideal}}  = 2^n$. \red{Nevertheless the value of the witness increases with $n$} (and hence the number of states in the Hilbert space) as expected.
More specifically, as we increase the number $n$ of qubits, the value of the corresponding quantum witness not only increases but is always larger than the maximum value with qubit number ($n-1$). 
This confirms that in practice the quantum witness can function as a dimensionality witness. We note that several other approaches to witnessing dimensionality, \red{using different types of temporal correlations},  were recently implemented \cite{Strikis18,spee2018genuine}.

\begin{table}
\centering  
\begin{tabular}{|c|c|c|c|c|} \hline
$n$ & $2$ & $3$ & $4$ & $6$  \\ \hline
$D_{\text{Ideal}}$ & $4$ & $8$ & $16$ & $64$  \\ \hline
$W_{\text{max}}$ & $0.75$ & $0.875$ & $0.937$ & $0.984$ \\ \hline
$W_{\text{Exp}}$ & $0.746\pm 0.005$ & $0.857\pm 0.004$ & $0.902 \pm 0.003$ & $0.940\pm 0.003$\\ \hline
$D_{\text{Exp}}$ & $\ge 3$ & $\ge 6$ & $\ge 10$ & $\ge 16$\\ \hline
\end{tabular}
\caption{Table of the quantum witness for a product of superposition states. Here, $D_{\text{Ideal}}=2^n$ is the ideal dimension of the system with qubit number $n$. The corresponding ideal value of the quantum witness is $W_{\text{max}}=1-1/D_{\text{Ideal}}$. Here, $W_{\text{Exp}}$ is the value of the quantum witness obtaining from the IBM Q$14$ Melbourne with the estimating dimension \red{$D_{\text{Exp}}=\lfloor (1-W_{\text{Exp}})^{-1}\rfloor$, where $\lfloor Y \rfloor$ is the integer of the number $Y$.}}
\label{table_superposition states}
\end{table}

\section{Discussion}
By taking advantage of the programmable nature of the IBM quantum experience, our results have shown how the violation of a Leggett-Garg inequality, in the form of a ``quantum witness'', changes  as we increase the number of qubits contributing to a highly entangled state. This allows us to see directly how the system becomes more macrorealistic as we increase the macroscopicity.

For $n=2$, we observed a violation of the quantum witness for $\theta=\pi/8,$ $2\pi/8$, $3\pi/8$, and $\pi/2$, and for $n=4$ for $\theta=\pi/2$. Thus, we can claim that when manipulating and observing two-qubits in the \rm{IBM Q$5$ Tenerife} device, and four qubits in the $14$-qubit processor \rm{IBM Q$14$ Melbourne} used for this experiments, the results must be described with a non-macrorealistic theory.
On the other hand, we found that six qubits, prepared in a GHZ state, did not violate the witness beyond a measurement invasiveness test, and thus these observations can, in principle, be described with macrorealistic theories. 
As the capabilities of the IBM quantum experience improve (e.g., when ancilla qubits are not required for the intermediate measurements), error correction, and error mitigation techniques are employed, the boundary between quantum theory and potential clumsy-macrorealistic theories could be tested with a much larger number of qubits.



The classical invasiveness, or clumisiness, we observed in the data (e.g., clearly exemplified by the non-zero quantum witness value at $\theta=0$ for $n=2$) can be explained by our ``minimal" Lindblad master equation noise model, where the infidelity of the CNOT operations used in the intermediate measurements causes changes in the state of the qubits. Moreover, our minimal model can also explain the suppression of the witness violation due to dephasing and energy relaxation.

To complement our primary results, instead of preparing entangled states, we also tested a product of superposition states, which has a low disconnectivity. We found that, as expected for such a state, the maximal violation increases with the number of qubits, and hence the dimensionality. In addition, the influence of noise on these results is substantially \red{smaller} than the GHZ-state based test. This is because single-qubit coherence tends to be less susceptible to noise than GHZ states, and because of the lower total circuit depth.

Finally, it is important to note that recent work has shown non-negligible non-Markovian effects in the IBM quantum experience~\cite{Pokharel2018}. This can introduce a secondary loophole in the LGI due to the non-instantaneous nature of the measurements we perform at time $t_1$. For example, in the IBM quantum experience, the measurements at time $t_1$ take about $0.4$ and $0.9$ $\mu \text{s}$ for the $5$ and $14$  qubit devices, respectively. This long-time scale appears because of the CNOT operation between primary and ancilla qubits needed for our intermediate measurement. Recent works suggest that non-Markovian effects are important on timescales of $\simeq 5$ $\mu \text{s}$~\cite{Pokharel2018}. Thus, differences in environment evolution on the timescale of our intermediate measurements may cause differences in the outcomes in the two contributions to the witness (i.e., differences to the final probability distributions between when the measurement is performed and when it is not). Like with clumsy measurements, because of this non-instantaneous measurement time, the origin of violations in this test due to breakdown of macroealism, or due to non-Markovian environmental influences, cannot be delineated. Our measurement invasiveness test may compensate for this to some degree, but further work is needed to take into account this potential loophole with such a test.
Alternatively, the non-Markovian effect can be diminished by using faster measurements, should such become available (either via faster CNOT operations, or the availability of direct measurements on the primary qubits at intermediate times).

In Appendix~\ref{Sec_prepare_and_measure} we consider an alternative approach to implement the witness which removes the need to use ancilla qubits, and hence reduce the circuit depth. From a simple inspection of the definition of the quantum witness, one can see that we can, instead of directly measuring the two-time correlation functions by using the ancilla qubits, first run an experiment where the probabilities $p_{t_1}^M(i)$ are collected. Then we run another experiment where one deterministically prepares the system in the state $i$, and measures $p_{t_2,t_1}(j|i)$. This scenario, which we call `prepare-and-measure', replaces the non-invasive measurement assumption with an ideal-state preparation and a more explicit non-Markovian evolution assumption~(see~\cite{CheMing12} and~\cite{Knee18}).


Overall, our results suggest that the current iteration of the IBM quantum experience tends towards clumsy-macrorealistic behavior for more than four qubits. This is inevitably also a function of the resulting circuit depth~\cite{Preskill2018} (i.e., overall run-time) on which the witness can be tested, which increases as the number of qubits is increased. A significant contribution to the circuit depth arises from the ancilla-based measurements, thus future improvements to the IBM quantum experience which allow multiple measurements on a single qubit may significantly reduce this circuit depth.

Finally, we point out that since a CNOT gate is its own inverse, one can reinterpret the combination of the quantum witness, and our choice of circuit, as a test of a \red{classical} circuit identity under the conditions of macrorealism. In other words, we tested whether CNOT$^2=\openone$ still holds under the condition of an \red{intermediate} projection onto a classical basis between the two CNOT gates. 
\red{Under quantum mechanics, of course, such relations are violated.  Thus,
we arrive at a different perspective on quantum witness tests, namely
that they can be viewed as tests of reversible classical circuit
identities under intermediate measurements.}

\section*{Data availability}
All data supporting the findings of this study including cat-states, measurement invasiveness, and product of superposition states have been deposited in creative commons with the DOI: 10.25405/data.ncl.9994739. Code for analysing the simulation and IBM qiskit is available on https://github.com/huan-yu-20/IBM-cat-state.

\begin{acknowledgments}
We acknowledge the NTU-IBM Q Hub (Giant: MOST 107-2627-E-002-001-MY3) and the IBM quantum experience for providing us a platform to implement the experiment. The views expressed are those of the authors and do not reflect the official policy or position of IBM or the IBM Quantum Experience team. The authors acknowledge fruitful discussions with George Knee, Jos\'e Gonz\'alez Alonso, Justin Dressel, Hong-Bin Chen, Jhen-Dong Lin, Yi-Te Huang and Kate Brown. H.-Y.K. acknowledges the support of the Graduate Student Study Abroad Program and Ministry of Science and Technology, Taiwan (Grant No. MOST 107-2917-I-006-002, and 108-2811-M-006-536 respectively). N.L. acknowledges partial support from JST PRESTO through Grant No. JPMJPR18GC.
This work is supported partially by the National Center for Theoretical Sciences and Ministry of Science and Technology, Taiwan, Grants No. MOST 107-2628-M-006-002-MY3, MOST 107-2811-M-006-017, and MOST 107-2627-E-006-001, and the Army Research Office (under Grant No. W911NF-19-1-0081). C. E. acknowledges support from the EPSRC grant EP/P034012/1. 
F.N. is supported in part by: NTT Research,
Army Research Office (ARO) (Grant No. W911NF-18-1-0358),
Japan Science and Technology Agency (JST) (via Q-LEAP and the CREST Grant No. JPMJCR1676), Japan Society for the Promotion of Science (JSPS) (via the KAKENHI Grant No. JP20H00134 and the JSPS-RFBR Grant No. JPJSBP120194828), and the Grant No. FQXi-IAF19-06 from the Foundational Questions Institute Fund (FQXi), a donor advised fund of the Silicon Valley Community Foundation.

\end{acknowledgments}

\section*{CONTRIBUTIONS:}
H.-Y. K,  N. L., and C. E. conceived the project and performed the theoretical analysis. H.-Y. K. and F.-R. J. ran the experiment and analysed the data. 
N. L., C. E., Y.-N. C. and F. N. supervised the research. All authors discussed the results and contributed to writing the manuscript.

\section*{Competing interests:} The authors declare no competing interests.


%


\clearpage
\appendix

\section{Modifying the quantum witness for clumsy measurements}\label{App_measurement_invasiveness_test}

In this section we explicitly derive Eq.~\eqref{Eq_disturbance_2} in the main text.  Inserting Eq.~\eqref{Eq_clumsy_measurement} into the definition of the witness, one finds
\beq
  W &=& \left| \sum_{i,k} p_{t_2,t_1}(j|k) \epsilon^{M}(k|i)p_{t_1}(i) - \sum_{i} p_{t_2,t_1}(j|i)p_{t_1}(i)\right|
  \nonumber\\
  &\leq & \max_{i}
  \left|\sum_{k} p_{t_2,t_1}(j|k) \epsilon^{M}(k|i)- p_{t_2,t_1}(j|i)\right|,
\label{george_1}
\eeq
where the maximum over states $i$ at time ${t_1}$ comes from the upper bound on a convex combination. This we can rewrite as 
\beq
  W 
  &\leq & \max_{i}
  \left|
    p_{t_2,t_1}(j|i) \left[1-\epsilon^{M}(i|i)\right]
    -\sum_{k\neq i} p_{t_2,t_1}(j|k) \epsilon^{M}(k|i) 
  \right|.
  \nonumber
\eeq
Since $p_{t_2,t_1}(j |i) \le 1$, $\sum_{k}\epsilon^{M}(k|i)=1$, and the remaining terms are positive, we can bound the right hand side further as 
\beq
  W 
  &\leq & 
  \max_{i}
  \left|
    1-\epsilon^{M}(i|i)
  \right|
  .
  \label{george_2}
\eeq
Thus we obtain an upper bound for the witness under the assumption of a \textit{macroscopically-invasive} nature of the intermediate measurements. This bound assumes nothing about the evolution from $t_1$ to $t_2$.

The bound in Eq.~\eqref{george_2} can be said to be a weaker bound than that in Eq.~\eqref{george_1} but is more experimentally efficient because we do not need to consider the effect of potential arbitrary evolution between $t_1$ and $t_2$.

We note that Eq.~\eqref{george_1} alone is equivalent to the test employed in~\cite{Knee18}. One just needs to sum up the outcomes $j$ in Eq.~\eqref{george_1} for the multi-outcome scenario considered in that work. Our additional derivation of a weaker bound in Eq.~\eqref{george_2} can be similarly generalized to multiple final outcome measurements.  This method is also related to the ``adroit" measurement test proposed in~\cite{Wilde2011,Huffman17}, when one assumes a particular intermediate measurement and that the states before that measurement are macrorealistic. Our bound is not as strong as the adroit one, but is easier to implement for the many-qubit situation we explore in this work.


\section{Quantum circuits: Direct-measure scenario}\label{Appendix quantum circuit direct measure}

From the initial state $\ket{0}^{\otimes n}$, ``cat states" at time $t_1$ can be obtained by performing the unitary transformation $U\ket{0}^{\otimes n}=\ket{\phi(n,\theta)}=\cos (\frac{\theta}{2})\ket{0}^{\otimes n}+\sin (\frac{\theta}{2})\ket{1}^{\otimes n}$. In the IBM quantum experience, we implemented the unitary $U$ by applying the $U_3(0,0,\theta)$ gate in Eq.~\eqref{Eq_U3} on the first qubit, and subsequently performing $n-1$ CNOT gates between the first qubits and all others. Therefore, the unitary $U=C^{Q_{n-2}}_{Q_{n-1}}......C^{Q_1}_{Q_2}C^{Q_0}_{Q_1}U_3^{Q_0}$ with the superscript $Q_0$ of a single quantum gate $U_3$ representing the operation acting on the qubit $Q_0$. 
Here, the super and sub scripts of a CNOT operation represent the control and target qubits, respectively. The inverse operation $U^{\dagger}$ is applied after time $t_1$ and it is given by reverse the gate implementation above.
We note that if one were to directly implement the circuit without `barriers' on the IBM quantum experience it would be automatically `optimized' to be an identity operation. In Fig.~\ref{Fig_circuits} (a), we present an explicit example of a two-qubit cat state.

Now, we can introduce how to perform the intermediate measurement at time $t_1$ and obtain the two-time correlation function. Since the IBM quantum experience only allows at most one measurement operation on any given qubit, we have to perform a CNOT gate on each measured qubit and an ancilla qubit.
Here, the ancilla and measured qubits are respectively the target and control qubits~[see Fig.~\ref{Fig_circuits} (b) and Ref.~\cite{NielsenBook}].
The measurement results on the ancilla qubit refer to the outcomes $i$ and leave behind the corresponding post-measurement states $\ket{\gamma}_i$. 
After the measurement at time $t_1$, we apply the $U^{\dagger}$ on the post-measurement state. We denote this approach as a direct-measure scenario. 

For instance, if the target and control qubits are respectively $\ket{0}$ and $\alpha\ket{0}+\beta\ket{1}$, with $|\alpha|^2+|\beta|^2=1$, the state after the CNOT operation is $\ket{\kappa}=\alpha\ket{00}+\beta\ket{11}$. Now we perform a measurement on the target qubit in the computational basis. Following Born's rule, we have 
\begin{equation}
\gamma_{i}=\text{Tr}_{\text{Target}}(\openone\otimes\ket{i}\bra{i}\rho),
\end{equation}
where $\rho=\ket{\kappa}\bra{\kappa}$ is the state at time $t_1$, $\ket{i}\bra{i}$ is a projector onto the computational basis, and $\gamma_{i}=\ket{\gamma}_i\bra{\gamma}$ is the remained state with the corresponding outcome $i$.

The second measurement with outcome $j$ at time $t_2$ can be implemented, without the need for ancillas. From this, the IBM quantum experience can return the result $ p_{t_2}^{M}(j)$. Finally, we note that while \rm{IBM Q$14$ Melbourne} has $14$ qubits, one cannot perform CNOT gates between arbitrary qubits because the direction of a CNOT gate is limited by the physical processor design~(see the physical structures in \cite{IBMQ}), limiting us to $6$ qubit in our cat state, and $6$ ancilla qubits.
We note that in the current IBM quantum experience, all of the qubits are measured in the end regardless of whether the measurement gates are actually implemented in the quantum circuit. After measuring all of the qubits, post-processing of the resulting data is applied according to the measurement gates one has chosen.

\section{The measurement invasiveness of the the other states}\label{Appendix_invasiness_four}
In this section, we present the values of the measurement invasiveness of the single, two four-qubit states in Tables~\ref{table_disturbance_two_three} and~\ref{table_disturbance_four}. We prepare all `macrorealistic' states $i$ in the computational basis to test the invasiveness of the intermediate measurement at time $t_1$.

In addition, it is important to note that the uncertainties given for the average values of the measurement invasiveness test represent the variance across $25$ different experiments (each individually consisting of 8192 runs) performed on different days, and thus reflect the variance in various properties of the IBM quantum experience across these long time scales~\cite{Alsina16}, and are thus different from the ones in the rest of the paper.

\begin{table}[H]
\centering  
\begin{tabular}{|c|c|c|c|} \hline
$\left\{i\right\}$ & $\II_{\text{Max}}(\left\{i\right\})$ & $\II_{\text{Ave}}(\left\{i\right\})$& $\II_{\text{Sim}}(\left\{i\right\})$  \\ \hline
$\left\{0\right\}$&$0.023\pm 0.004$ & $0.016\pm 0.003$ & $0.031$\\ \hline
$\left\{1\right\}$ &$0.024\pm 0.004$ &$0.014\pm 0.004$ & $0.011$\\ \hline \hline
$\left\{00\right\}$&$0.077\pm 0.008$ & $0.068\pm 0.006$ & $0.061$\\ \hline
$\left\{11\right\}$ &$0.051\pm 0.005$ &$0.033\pm 0.009$ & $0.021$\\ \hline
$\left\{10\right\}$ &$0.033\pm 0.004$ &$0.020\pm 0.006$ & $0.019$\\ \hline
$\left\{01\right\}$ &$0.040 \pm 0.004$ &$0.030\pm 0.005$ & $0.019$\\ \hline
\end{tabular}
\caption{Table of the measurement invasiveness parameters for single, two-qubit systems, respectively. Here we perform the $25$ experiments, across multiple days, to take into account variability in the IBM quantum system parameters. Each collation consists of $8192$ runs. The maximal and average values over experiments of the measurement invasiveness are obtained from \rm{the IBM Q$5$ Tenerife} for the single and two-qubit systems.
 The simulation results, using the noise model described in the main text, are also presented.
}

\label{table_disturbance_two_three}
\end{table} 

\begin{table}[H]
\centering  
\begin{tabular}{|c|c|c|c|c|} \hline
$\left\{i\right\}$ & $\II_{\text{Max}}(\left\{i\right\})$ &  $\II_{\text{Ave}}(\left\{i\right\})$  \\ \hline
$\left\{0000\right\}$&$0.146\pm 0.005$ & $0.106\pm 0.019$\\ \hline
$\left\{0001\right\}$&$0.115\pm 0.005$ & $0.075\pm 0.029$ \\ \hline
$\left\{0010\right\}$&$0.123\pm 0.006$ & $0.104\pm 0.011$ \\ \hline
$\left\{0100\right\}$&$0.136\pm 0.006$ & $0.093\pm 0.026$ \\ \hline
$\left\{1000\right\}$&$0.100\pm 0.006$ & $0.078\pm 0.015$ \\ \hline
$\left\{0011\right\}$&$0.091\pm 0.006$ & $0.067\pm 0.013$ \\ \hline
$\left\{0101\right\}$&$0.124\pm 0.006$ & $0.079\pm 0.031$ \\ \hline
$\left\{1001\right\}$&$0.097\pm 0.006$ & $0.061\pm 0.023$ \\ \hline
$\left\{0110\right\}$&$0.121\pm 0.006$ & $0.103\pm 0.012$ \\ \hline
$\left\{1010\right\}$&$0.118\pm 0.006$ & $0.086\pm 0.022$ \\ \hline
$\left\{1100\right\}$&$0.122\pm 0.006$ & $0.080\pm 0.020$ \\ \hline
$\left\{1110\right\}$&$0.109\pm 0.007$ & $0.078\pm 0.022$ \\ \hline
$\left\{1101\right\}$&$0.068\pm 0.007$ & $0.052\pm 0.010$ \\ \hline
$\left\{1011\right\}$&$0.077\pm 0.006$ & $0.046\pm 0.018$ \\ \hline
$\left\{0111\right\}$&$0.095\pm 0.007$ & $0.057\pm 0.021$ \\ \hline
$\left\{1111\right\}$&$0.070\pm 0.007$ & $0.035\pm 0.018$ \\ \hline
\end{tabular}
\caption{Table of the measurement invasiveness parameters for four-qubit systems.}
\label{table_disturbance_four}
\end{table}

\section{Quantum circuits: Prepare-and-measure scenario}\label{Sec_prepare_and_measure}
An alternative approach (which can in principle allow for a larger number of measured qubits since no ancilla qubits are needed) relies on trading the measurement at time $t_1$ with ideal state preparation. In this new scenario, the first circuit is performed with a unitary transformation $U$ before the measurements at time $t_1$. The IBM quantum experience returns the probability distribution $p_{t_1}(i)$ with outcomes $i$. According to the probability distribution $p_{t_1}(i)$, we then prepare a new circuit with an initial state in the eigenstates $\ket{i}$. The $U^{\dagger}$ operation is then performed before the measurements at time $t_2$ on the system. The results from the IBM quantum experience represent the conditional probability distributions $p_{t_2,t_1}(j|i)$. Here, only the outcome $j=0$ is used to analyse the quantum witness in Eq.~\eqref{Eq_quantum_witness}.

We prepare all possible eigenstates $\ket{i}$ for $n=2$ and $4$ qubits systems. For the $6$ qubit case, we only prepare the eigenstates $\ket{i}$ if $p_{t_1}(i)\geq 10^{-3}$, which is chosen to be much smaller than the ideal outcome of, e.g., $p^M_{t_1}(0)=0.5$ (note that the error induced in the witness due to omission of these small terms can in principle be of the same order as the uncertainty in the experimental data we show later; but given that the observed violation is already lower than the measurement invasiveness, this error does not cause a false witness).
Finally, we note that there are at most $(i+1)$ quantum circuits in this scenario with $i$ being the total number of the states we need to prepare. However, there are only two experimental circuits needed to collate the corresponding statistical data $\sum_i p_{t_1}(i_{t_1})p_{t_2,t_1}(j|i)$, and $p_{t_2}(j)$ in the direct-measure scenario.
Therefore, the prepare-and-measure scenario is not efficient as the number of qubits increases because the number of quantum circuits we need to collate all possible correlations increases with the number of outcomes $i$. 


As with the direct-measure scenario, which suffers from a ``clumsiness loophole'' arising from the noninvasive measurement assumption, the prepare-and-measure scenario can similarly suffer from a clumsiness loophole related to non-ideal state preparation which lead a non-zero value for $\theta=0$ in our experiment. Moreover, in principle, non-Markovian effects also lead to a false positive violation of the quantum witness. For instance, if the history from time $t_0$ to $t_1$ influences the evolution from time $t_1$ to $t_2$, this may also lead to differences in the probability distributions $p_{t_2}(j)$ and $p_{t_2}^{M}(j)$~\cite{Knee18,CheMing12}.

\begin{figure*}[tbp]
\includegraphics[width=2\columnwidth]{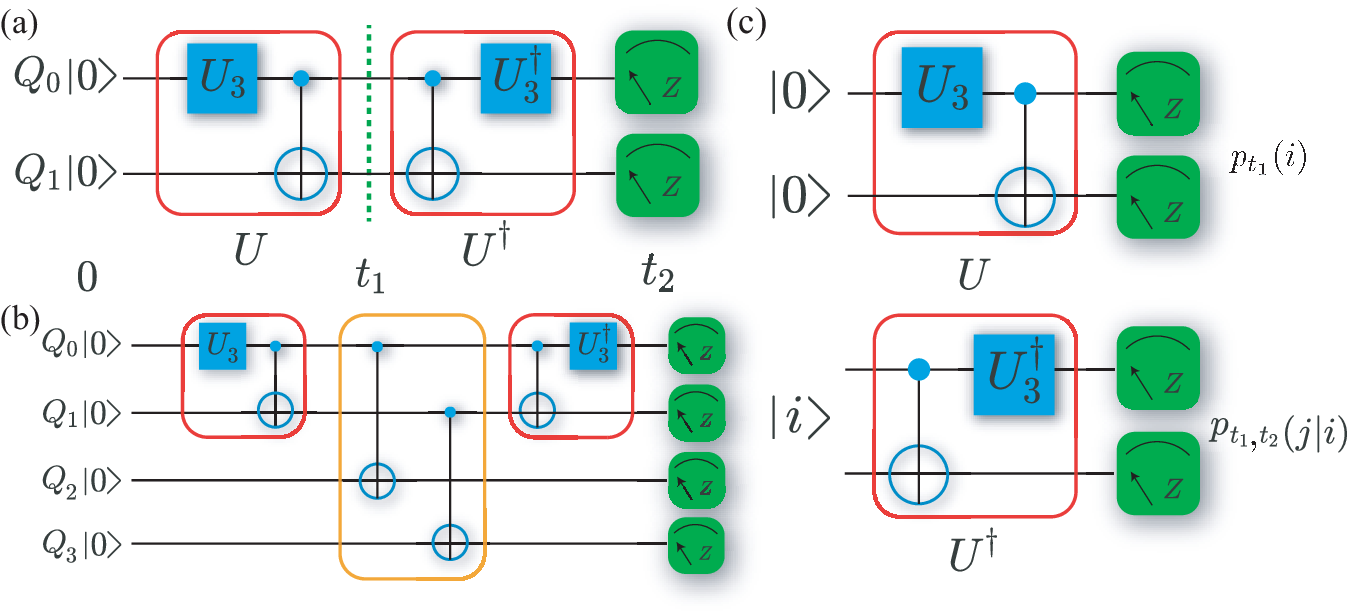}
\caption{Example of schematic quantum circuits for $n=2$.
(a) is for measuring $p_{t_2}(j)$. In the IBM quantum experience, the qubits denoted by $Q_i$ for $i=0$ and $1$ are initially prepared in $\ket{0}$. The left and right red areas respectively represent the unitary transformations $U$ and $U^{\dagger}$ which can be decomposed by $U_3(0,0,\theta)$ ($U_3$ in short) and a series of CNOT operations. In the beginning, $U_3$ is performed on $Q_0$, followed by a CNOT gate on the control $Q_0$ and the target qubits $Q_1$. The green dots represent the barrier between $U$ and $U^{\dagger}$ to avoid the automatic optimization. The $U^{\dagger}$ is performed after the barrier. In the end, the measurements on the computational basis are performed such that the value $p_{t_2}(j)$ is obtained.~(b) shows the quantum circuit for measuring $\sum_i p_{t_1}(i)p_{t_2,t_1}(j|i)$ in the direct-measure scenario. Since the IBM quantum experience cannot measure the same qubit twice, the intermediate measurement at time $t_1$ can be implement by the CNOT operation with the ancilla qubit $Q_2$ and $Q_3$. We use the yellow box to represent the intermediate measurement. Here, the ancilla qubits are initially in $\ket{0}$. Since we only consider the projective measurement onto the computational basis, one can implement the CNOT operation to transfer the classical information of the state to the ancilla qubit. The measurement operations on the ancilla qubits $Q_2$ and $Q_3$ remain in the post-measurement state $\ket{\gamma_i}$ with outcomes $i$. Finally, with the measurement on the qubits $Q_1$ and $Q_2$, the quantum circuit returns the result $\sum_i p_{t_1}(i)p_{t_2,t_1}(j|i)$.~(c) shows the quantum circuits for, respectively, measuring $p_{t_1}(i)$ and $p_{t_2,t_1}(j|i)$ in the prepare-and-measured scenario. The unitary transformation $U$ is performed on the state $\ket{0}$, followed by measurement operations with outcome $i$ at time $t_1$. In the second experiments, the eigenstates $\ket{i}$ are prepared according to the probability $p_{t_1}(i)$, followed by the inverse unitary transformation $U^{\dagger}$. The measurement results are the probability with outcome $j$ conditional on $i$.
}
\label{Fig_circuits} %
\end{figure*}

In the end of the section, we present a schematic example of the quantum circuit for the two-qubit case [see Fig.~\ref{Fig_circuits} (c))].
The initial state of the total system on the IBM quantum experience is $\ket{0}^{\otimes 2}$. The state becomes a ``cat state" in Eq.~\eqref{Eq_cat_state} by applying a unitary transformation $U$. Instead of evolving back to the state $\ket{0}^{\otimes 2}$, we measure the cat states at time $t_1$ to obtain the probability $p_{t_1}(i)$ [see the top half of the Fig.~\ref{Fig_circuits} (c)]. After the first experiment, we prepare a quantum state $\ket{i}$, which is the eigenstate of the corresponding outcomes $i$, and perform a $U^{\dagger}$ operation. Finally, the measurement operation is performed to obtain the probability $p_{t_2,t_1}(j|i)$ [see the bottom of the Fig.~\ref{Fig_circuits} (c)]. One can easily expand the two-qubit system to a GHZ one.

\begin{table}[!tbp]
\centering  
\begin{tabular}{c}
\text{Quantum witness for cat state with $n=2$}
\end{tabular}
\begin{tabular}{|c|c|c|} \hline
& $W_{PM}$& $W_{DM}$\\ \hline
$\theta=0$ & $0.061\pm 0.006$  &$0.058\pm 0.007$\\ \hline
$\theta=\pi/8$ & $0.112\pm 0.006$ &$0.102\pm 0.007$\\ \hline
$\theta=2\pi/8$ & $0.261\pm 0.006$  &$0.250\pm 0.008$\\ \hline
$\theta=3\pi/8$ & $0.430\pm 0.006$  &$0.376\pm 0.008$\\ \hline
$\theta=4\pi/8$ & $0.454\pm 0.006$  &$0.453\pm 0.008$\\ \hline
\end{tabular}
\caption{Table of quantum witness of cat states with $n=2$. The $W_{PM}$ and $W_{DM}$ are quantum witness obtained by prepare-and-measure and direct-measure scenarios, respectively. We note that the measurement invasiveness is $0.077\pm 0.008$.}
\label{Table_small_cat}
\end{table}

In general, the prepare-and-measure scenario can also test for qubit number $n> 6$. However, we do not do this cumbersome procedure because the direct-measure results shows that for the $n=6$ case the system is already classified as macrorealistic.

\begin{table}[H]
\centering  
\begin{tabular}{c}
\text{Quantum witness for cat states with $n=4$ and $6$}
\end{tabular}
\begin{tabular}{|c|c|c|} \hline
& $W_{PM}$& $W_{DM}$\\ \hline
$n=4$ & $0.251\pm 0.007$  &$0.245\pm 0.008$\\ \hline
$n=6$ & $0.183\pm 0.067$ &$0.181\pm 0.007$\\ \hline
\end{tabular}
\caption{Table of quantum witness of cat states with $n=4$ and $n=6$. The $W_{PM}$ and $W_{DM}$ are quantum witness obtained by prepare-and-measure and direct-measure scenarios, respectively.}
\label{Table_large_cat}
\end{table}

Interestingly, the witness values from the prepare-and-measure scenario are almost all slightly higher than the direct-measure ones [see Table~\ref{Table_small_cat} and \ref{Table_large_cat}]. From the circuit-implementation point of view, the prepare-and-measure scenario significantly reduces the number of CNOT gates, which take almost four times longer than the $U_3$ gates. Therefore, the prepare-and-measure scenario effectively reduces the overall effect of noise on the witness and has a much lower circuit depth. However even the prepare-and-measure scenario does not produce a violation for six qubits. 




\section{Figure legends}
Fig.~1:Schematic setup. We prepare $n$ qubits on the state $\ket{0}^{\otimes n}$ (blue) at time $t_0$. A unitary $U$ transfers the system into the entangled cat state $\ket{\phi(n,\theta)}=\cos{\frac{\theta}{2}}\ket{0}^{\otimes n}+\sin{\frac{\theta}{2}}\ket{1}^{\otimes n}$ (red) at time $t_1$. Then, an inverse unitary $U^{\dagger}$ is performed to the entangled system, such that the system returns back to the state $\ket{0}^{\otimes n}$ at time $t_2$. The outcomes $i$ and $j$ are obtained at $t_1$ and $t_2$, respectively.

Fig.~2: The value of the quantum witness of $n=2$ cat states. The circuit is designed to produce the state $\ket{\phi(2,\theta)}=\cos{\frac{\theta}{2}}\ket{0}^{\otimes 2}+\sin{\frac{\theta}{2}}\ket{1}^{\otimes 2}$ at an intermediate time. The experimental results obtained by the IBM quantum experience are shown by blue diamonds. The theoretical results, with and without noise simulation, are shown by red dashed and black solid curves, respectively. Obviously, the quantum witness increases with the parameter $\theta$, but also shows a residual violation due to the macroscopically-invasive measurements backaction and gate error at $\theta =0$.
We simulate the influence of decoherence and gate infidelities by Lindblad-form master equations Eq.~\eqref{Eq_master_D} and Eq.~\eqref{gamerr}. The coefficients of relaxation time $T_1=46~\mu$s, dephasing time $T_2=13.5~\mu$s, and gate-error coefficient $\gamma_{\text{Errors}}=8.5\times 10^{-2}~\mu \text{s}^{-1}$ are determined by approximately fitting the experimental results. The gray and orange shaded areas at the bottom are the clumsy-macrorealistic regimes determined by the maximal and average invasiveness tests in Table~\ref{table_disturbance_two}.
Note that the invasiveness test including the standard deviation does not depend on $\theta$. The experimental uncertainties are derived from the multinomial distribution and error propagation except for the average disturbance case which is the variance across $25$ repeated experiments.

Fig.~3: Example of schematic quantum circuits for $n=2$.
(a) is for measuring $p_{t_2}(j)$. In the IBM quantum experience, the qubits denoted by $Q_i$ for $i=0$ and $1$ are initially prepared in $\ket{0}$. The left and right red areas respectively represent the unitary transformations $U$ and $U^{\dagger}$ which can be decomposed by $U_3(0,0,\theta)$ ($U_3$ in short) and a series of CNOT operations. In the beginning, $U_3$ is performed on $Q_0$, followed by a CNOT gate on the control $Q_0$ and the target qubits $Q_1$. The green dots represent the barrier between $U$ and $U^{\dagger}$ to avoid the automatic optimization. The $U^{\dagger}$ is performed after the barrier. In the end, the measurements on the computational basis are performed such that the value $p_{t_2}(j)$ is obtained.~(b) shows the quantum circuit for measuring $\sum_i p_{t_1}(i)p_{t_2,t_1}(j|i)$ in the direct-measure scenario. Since the IBM quantum experience cannot measure the same qubit twice, the intermediate measurement at time $t_1$ can be implement by the CNOT operation with the ancilla qubit $Q_2$ and $Q_3$. We use the yellow box to represent the intermediate measurement. Here, the ancilla qubits are initially in $\ket{0}$. Since we only consider the projective measurement onto the computational basis, one can implement the CNOT operation to transfer the classical information of the state to the ancilla qubit. The measurement operations on the ancilla qubits $Q_2$ and $Q_3$ remain in the post-measurement state $\ket{\gamma_i}$ with outcomes $i$. Finally, with the measurement on the qubits $Q_1$ and $Q_2$, the quantum circuit returns the result $\sum_i p_{t_1}(i)p_{t_2,t_1}(j|i)$.~(c) shows the quantum circuits for, respectively, measuring $p_{t_1}(i)$ and $p_{t_2,t_1}(j|i)$ in the prepare-and-measured scenario. The unitary transformation $U$ is performed on the state $\ket{0}$, followed by measurement operations with outcome $i$ at time $t_1$. In the second experiments, the eigenstates $\ket{i}$ are prepared according to the probability $p_{t_1}(i)$, followed by the inverse unitary transformation $U^{\dagger}$. The measurement results are the probability with outcome $j$ conditional on $i$.

\end{document}